\documentclass[prb,showpacs,twocolumn,amsmath,amssymb,floatfix,superscriptaddress]{revtex4}
\usepackage{graphicx}
\usepackage{bm}
\usepackage{float}
\usepackage{psfrag}

\begin{document}

\title{Lower limit on the achievable temperature in resonator-based sideband cooling}

\author{M.~Grajcar}
\affiliation{Frontier Research System, The Institute of Physical
and Chemical Research (RIKEN), Wako-shi 351-0198, Japan}
\affiliation {Department of Experimental Physics, Comenius
University, SK-84248 Bratislava, Slovakia.}

\author{S. Ashhab}
\affiliation{Frontier Research System, The Institute of Physical
and Chemical Research (RIKEN), Wako-shi 351-0198, Japan}
\affiliation{Physics Department, The University of Michigan, Ann
Arbor, Michigan 48109-1040, USA}

\author{J.R. Johansson}
\affiliation{Frontier Research System, The Institute of Physical
and Chemical Research (RIKEN), Wako-shi 351-0198, Japan}

\author{Franco Nori}
\affiliation{Frontier Research System, The Institute of Physical
and Chemical Research (RIKEN), Wako-shi 351-0198, Japan}
\affiliation{Center for Theoretical Physics, Physics Department,
Applied Physics Program, Center for the Study of Complex Systems,
The University of Michigan, Ann Arbor, Michigan 48109-1040, USA}
\affiliation{CREST, Japan Science and Technology Agency (JST), Kawaguchi,
Saitama 332-0012, Japan}

\date{\today}

\begin{abstract}
A resonator with eigenfrequency $\omega_r$ can be effectively used
as a cooler for another linear oscillator with a much smaller
frequency $\omega_m\ll\omega_r$. A huge cooling effect, which
could be used to cool a mechanical oscillator below the energy of
quantum fluctuations, has been predicted by several authors.
However, here we show that there is a lower limit $T^*$ on the
achievable temperature, given by $T^*=T_m\omega_m/\omega_r$, that
was not considered in previous work and can be higher than the
quantum limit in realistic experimental realizations. We also
point out that the decay rate of the resonator, which previous
studies stress should be small, must be larger than the decay rate
of the cooled oscillator for effective cooling.
\end{abstract}

\pacs{85.85.+j, 45.80.+r}

\maketitle

\section{\label{sec:Introduction} Introduction}
Recently, a tremendous experimental effort has been devoted to the
task of cooling mechanical oscillators below the energy of quantum
fluctuations. In spite of many experimental improvements, the
quantum limit has not been
achieved.\cite{LaHaye04,Schwab05,Gigan06,Schliesser06,Brown-2007}
Several papers that propose cooling mechanisms using
electromagnetic (rf, microwave or light)
resonators\cite{wilson-rae04,Xue-2007,marquardt07,wilson-rae07} or
other cooling
mechanisms\cite{martin04,Hensinger05,Grajcar07,Quan-2006,You08,DykmanNote}
to fulfill this task have appeared recently. These papers predict
an enormous cooling effect. However, they do not explicitly state
that there is a lower limit on the achievable temperature,
associated with the ratio between the frequencies of the coolant
and cooled oscillators, that cannot be overcome and can play an
important role for realistic experimental realizations. Moreover,
some formulas that appear in the literature can give temperatures
below this limit, which will be described in more detail below.
This lower temperature limit can be important for the most
feasible designs using rf or microwave resonators.

The electromagnetic resonators can be easily implemented on-chip
beside a nano-mechanical oscillator and
kept at low temperatures. Such structures, also known as MEMS and
NEMS (micro and nanoelectromechanical systems), have already been
realized \cite{Huang03,Sazanova04, Peng06}, achieving high
frequencies in the GHz range ($\omega_m\sim 10$~GHz) but with
small quality factors $Q_m\sim10-500$. Nevertheless, they can be
used as sensitive elements for weak force-displacement
detection\cite{wei06, milburn-2007}
and they have been proposed as possible qubits.\cite{You05,Savelev06}
For such systems, the frequency of
the basic mode of the NEMS becomes comparable to the resonance
frequency of the electromagnetic resonator $\omega_r$. In this
case, the temperature limit proportional to $\omega_m/\omega_r$
which was negligible for optical-frequency coolers, can determine
the lowest achievable temperature $T^*$. The aim of this work is
to call the attention of experimentalists to this fundamental
limit which could help them design more effective cooling systems.

\section{\label{sec:Classical} Semi-Classical approach}
\begin{figure}
    \psfrag{oV}{$\tilde{V}_0$}
        \psfrag{Te/Tm}{$T^*_m/T_m$}
\includegraphics[width=7.0cm]{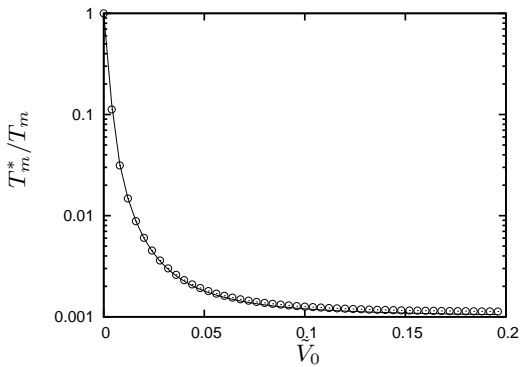}
\caption{The cooling factor $T^*_m/T_m$
as a function of the normalized pumping amplitude $\tilde{V}_0$
of the noiseless microwave source for $\tilde{\omega}_r=10^3$,
$\tilde{\Gamma}_m=10^{-5}$, $\tilde{\Gamma}_r=10^{-1}$ and
$k_BT_m=k_BT_r\gg\hbar\omega_{r,m}$ calculated numerically (circles) and from
Eqs.~(\ref{Eq:svv}),(\ref{Eq:sigma33}) (solid
line).}
\label{Fig:T_m}
\end{figure}

For the sake of simplicity, we will consider a RLC tank circuit
(the results can be applied to any electromagnetic resonator, such
as a transmission-line resonator, cavity, Fabry-Perot resonator,
etc.). A mechanical oscillator is coupled to the capacitor such
that the capacitance depends parametrically on the displacement of
the oscillator. Such a system was thoroughly analyzed in
Ref.~\onlinecite{Braginsky}, and we only briefly introduce the
equations of motion here. If the mechanical oscillator is a part
of one of the capacitor electrodes, the capacitance $C(x)\approx
C_0(1-x/d)$ depends on the displacement $x$ of the oscillator from
the equilibrium position, where $C_0=\epsilon S/d_0$ is the
capacitance at $x=0$ and $d$ is the renormalized distance between
the electrodes $d=d_0/\kappa$. Here $\kappa$ is the coupling
constant between the mechanical oscillator and the RLC circuit,
and it can be expressed as the ratio between the mechanical
oscillator capacitance $C_m$, which depends  on the oscillator
displacement, and the total capacitance $C_0$ (we consider the
case $C_m\ll C_0$), i.e. $\kappa=C_m/C_0$.  If the RLC tank
circuit is pumped by a microwave source $V_p=V_{p0}\cos\omega_pt$,
the voltage between the capacitor's electrodes is
$V_0=\omega_rV_{p0}/\Gamma_r$, and the Coulomb energy of the
capacitor depends on its capacitance which, in turn, depends on
the oscillator displacement. Thus, the electromagnetic resonator
and mechanical oscillator (i.e. cantilever) can be described by a system of
differential equations of two coupled damped linear oscillators
\begin{eqnarray}
\frac{d^2Q}{dt^2}&+&\Gamma_r\frac{dQ}{dt}+\omega^2_rQ\left(1-\frac{x(t)}{d}\right)=
\frac{V_p(t)+V_f(t)}{L}\ \label{Eq:LRCxa}\\
    \frac{d^2x}{dt^2}&+&\Gamma_m\frac{dx}{dt}+\omega_m^2x=
\frac{F_f(t)}{M}+\frac{Q^2(t)}{2MC_0d}
     \label{Eq:LRCxb}
\end{eqnarray}
where $\Gamma_{r,m}$ are damping rates, $\omega_{r,m}$ are angular
frequencies, $V_f$ is a fluctuating voltage across the capacitor,
$F_f$ is a fluctuating force acting on the mechanical oscillator
with mass $M$, and $Q(t)=q_p(t)+q_f(t)$ is total  charge charge on the capacitor.
The Eq.~(\ref{Eq:LRCxb}) is nonlinear but can be linearized keeping in mind that we are
interested to calculate charge fluctuations $q_f(t)$ which are much smaller than charge
oscillations $q_p(t)$ driving by coherent microwave source.
It is convinient to express $q_f(t)$ and $V_f(t)$ in terms of quadrature
amplitudes
\begin{eqnarray}
\nonumber
q_f(t)&=&q_c(t)\cos\omega_pt+q_s(t)\sin\omega_pt \ ,\\ \nonumber
V_f(t)&=&V_c(t)\cos\omega_pt+V_s(t)\sin\omega_pt\ ,\\ \nonumber
\end{eqnarray}
and rewrite Eqs.~(\ref{Eq:LRCxa}),(\ref{Eq:LRCxb}) in the dimensionless
variables
\begin{eqnarray}
\nonumber
\tilde{q}_{c,s}&\equiv& q_{c,s}/\sqrt{C_0\hbar\omega_r}\ ,\\ \nonumber
\tilde{x}&\equiv& x/\sqrt{\hbar\omega_m/M\omega^2_m}\ ,\\ \nonumber
\tau&\equiv&\omega_mt\ ,\\ \nonumber
\tilde{\omega}_{r,p}&\equiv&\omega_{r,p}/\omega_m\ ,\\ \nonumber
\tilde{\Gamma}_m&\equiv&\Gamma_m/\omega_m\ ,\\ \nonumber
\tilde{\Gamma}_r&\equiv&\Gamma_r/2\omega_m\ ,\\ \nonumber
\tilde{V}_{c,s}&\equiv&\tilde{\omega}_rV_{c,s}\sqrt{C_0/\hbar\omega_r}\ ,\\ \nonumber
\tilde{F}_f&\equiv&F_f/\sqrt{M\omega^2_m\hbar\omega_m}\ ,\\ \nonumber
\tilde{V}_0&\equiv&\tilde{\omega}_rV_0\sqrt{C_0/4M\omega_m\omega_rd^2}. \nonumber
\end{eqnarray}
Here $T_r$ and $T_m$ are the temperatures
of the electromagnetic resonator and mechanical oscillator,
respectively.
Using the slowly-varying-amplitude
approximation\cite{Braginsky} and considering Langevin fluctuating
forces caused by quantum noise,\cite{Gardiner,GardinerZoller}
\begin{eqnarray}
\nonumber
V_{c,s}(t)&=&\sqrt{L\Gamma_r\frac{\hbar\omega_r}{2}\coth\left(\frac{\hbar\omega_r}{2k_BT_r}\right)}\xi_{c,s}(t)\ ,\\ \nonumber
F_f(t)&=&\sqrt{M\Gamma_m\hbar\omega_m\coth\left(\frac{\hbar\omega_m}{2k_BT_m}\right)}\xi_m(t)\ ,
\end{eqnarray}
Eqs.~(\ref{Eq:LRCxa}),(\ref{Eq:LRCxb}) read
\begin{equation}
    \frac{d\tilde{q}}{d\tau}= -\tilde{A} \, \tilde{q}(\tau)+\tilde{F}(\tau)
 \label{Eq:difeqarr}
\end{equation}
where
\begin{equation}
    \tilde{A}=\left(\begin{array}{cccc}
        \tilde{\Gamma}_r & \tilde{\eta}    &     0      &  0         \\
        -\tilde{\eta}     & \tilde{\Gamma}_r &     0      &  -\tilde{V}_0  \\
        -\tilde{V}_0      &    0     &  \tilde{\Gamma}_m & 1 \\
                   0      &    0     &     -1      &  0         \\
       \end{array}\right)\ ,
    \label{Eq:A}
\end{equation}
\begin{equation}
    \tilde{F}(\tau)=\left(\begin{array}{c}
              \sqrt{\tilde{\Gamma}_r\coth (\hbar\omega_r/2k_BT_r)}\xi_c(\tau) \\
              \sqrt{\tilde{\Gamma}_r\coth (\hbar\omega_r/2k_BT_r)}\xi_s(\tau) \\
              \sqrt{\tilde{\Gamma}_m\coth (\hbar\omega_m/2k_BT_m)}\xi_m(\tau) \\
                   0
               \end{array}\right)\ ,
    \label{Eq:F}
\end{equation}
$\tilde{q}\equiv (\tilde{q}_c,\tilde{q}_s,\tilde{v},\tilde{x})$,
$\tilde{v}\equiv d\tilde{x}/d\tau$ and
$\tilde{\eta}=\tilde{\omega}_p-\tilde{\omega}_r$.
Here $T_r$ and
$T_m$ are the base temperatures of the resonator and mechanical
oscillator, respectively.
Thus, we have a system of coupled Langevin
equations\cite{Gardiner} which allow us to calculate the
stationary covariance matrix defined as $\sigma\equiv\langle
\tilde{q}\tilde{q}^T\rangle_s$ for $\tau\rightarrow\infty$. The diagonal
terms of the covariance matrix determine the mean squared values of the
vector components $\tilde{q}$. For example,
$\sigma_{\tilde{v}\tilde{v}}\equiv\langle \tilde{v}^2\rangle_s$ is
the normalized mean squared velocity of the mechanical oscillator.
The covariance matrix can be determined from the system of linear
equations
$$\tilde{A} \, \sigma+\sigma \tilde{A}^T=\tilde{B}\ ,$$
where
$\tilde{B}$ is a correlation matrix defined as
$\langle\tilde{F}_i(\tau)\tilde{F}_j(\tau')\rangle=\tilde{B}_{ij}\delta(\tau-\tau')$.
If the fluctuating forces are uncorrelated, i.e.
$\langle\xi_x(\tau)\xi_{x'}(\tau')\rangle=\delta_{x,x'}\delta(\tau-\tau')$
(here $x,x'$ stand for $c,s$ or $m$), $\tilde{B}$ takes the form
of a diagonal matrix with elements
\begin{equation}
    \tilde{B}_{ii}=\left(\begin{array}{c}
    \tilde{\Gamma}_r\coth(\hbar\omega_r/2k_BT_r)\ \\
    \tilde{\Gamma}_r\coth(\hbar\omega_r/2k_BT_r)\ \\
        \tilde{\Gamma}_m\coth(\hbar\omega_m/2k_BT_m)\ \\
     0
                  \end{array}\right)\ .
    \label{Eq:B}
\end{equation}
The mean value of energy of the mechanical oscillator fluctuations is
\begin{equation}
{\cal E}_m=\sigma_{\tilde{v}\tilde{v}} \, \hbar \, \omega_m.
    \label{Eq:Em}
\end{equation}
Now, one can easily calculate the effective temperature of the
mechanical oscillator from the definition relation for $T^*_m$
\begin{equation}
    \sigma_{\tilde{v}\tilde{v}}=\frac{1}{2}\coth\left(\frac{\hbar\omega_m}{2k_BT^*_m}\right)\ .
    \label{Eq:svv}
\end{equation}
As we will see later, the most appropriate parameters for cooling
purposes are $\tilde{\eta}=-1$,
$2\tilde{\Gamma}_m\tilde{\Gamma_r}\ll\tilde{V}_0\ll 1$. In this limit and for
$\Gamma_r<\omega_m$, $\Gamma_m\ll\Gamma_r$ the $\sigma_{\tilde{v}\tilde{v}}$ can
be expressed as
\begin{equation}
    \sigma_{\tilde{v}\tilde{v}}=\frac{1}{2}
    \coth\left(\frac{\hbar\omega_r}{2k_BT_r}\right)+
    \frac{\tilde{\Gamma}_r\tilde{\Gamma}_m}{\tilde{V}^2_0}
    \coth\left(\frac{\hbar\omega_m}{2k_BT_m}\right)
    \label{Eq:sigma33}
\end{equation}
Thus, the lowest  temperature of the mechanical oscillator is
limited by the first term if the second term is made negligibly
small by sideband cooling. As a matter of fact this term simply
shows that even in our semi-classical approach we cannot `cool'
the mechanical oscillator below the zero-point energy which is
consistent with the Heisenberg uncertainty principle. Indeed, it
follows from Eqs.(\ref{Eq:Em}) and (\ref{Eq:svv}) that the energy
saved in the mechanical oscillator is
\begin{equation}
    {\cal E}_m=\frac{\hbar\omega_m}{2}
    \coth\left(\frac{\hbar\omega_r}{2k_BT_r}\right)
    \label{Eq:Eml}
\end{equation}
In this limit,
the effective temperature $T^*_m$ of the mechanical oscillator takes the simple form
\begin{equation}
    T^*_m=\frac{\omega_m}{\omega_r}T_r
    \label{Eq:T_ml}
\end{equation}
The cooling factor $T^*_m/T_m$ as a
function of the normalized pumping amplitude $\tilde{V}_0$ is
shown in Fig.~\ref{Fig:T_m}.
Even though this result was derived within semi-classical physics, the
same limit can be obtained using the quantum approach, as we shall
show below.

Here we should emphasize that the temperature of the resonator
$T_r$ is usually much higher than the ambient temperature if the
resonator is heavily pumped by the microwave source. This is
caused by the phase noise of the microwave source, which is directly
proportional to the output power. Microwave sources are
characterized by the single sideband noise spectral
density\cite{Lee00}
\begin{equation}
L(\Delta\omega)=10\log\left(\frac{S_V}{U^2_{mw}}\right)\ ,
    \label{Eq:Lom}
\end{equation}
where $U^2_{mw}$ is the mean-square voltage of the microwave
source and $S_V$ is the spectral density of the voltage noise. The
best commercially available microwave sources achieve
$L(\Delta\omega)=-170$~dBc/Hz. The effective temperature $T_r$ of
the pumped resonator can be calculated as
\begin{equation}
    T_r=T_{r0}+\frac{\omega_r}{\Gamma_r}\frac{U^2_{mw}}{2k_BZ_r}10^{L(\Delta\omega)/10}
    \label{Eq:L}
\end{equation}
where $T_{r0}$ is the temperature of the resonator without pumping
and $Z_r=\sqrt{L/C}$ is the characteristic impedance of the
resonator. Now, both terms in Eq.~(\ref{Eq:sigma33}) depend on the pumping
power. The first one increases with pumping power while the second one decreases.
Since the highest cooling power is expected for\cite{Xue-2007}
$\tilde{\Gamma}_r<1$ the effective temperature of the
mechanical oscillator is higher than
\begin{equation}
T_{m0}\approx
\frac{U^2_{mw}}{2k_BZ_r}10^{L(\Delta\omega)/10}\ .
    \label{Eq:Tm0}
\end{equation}
Thus, for microwave resonators the first term in
Eq.~(\ref{Eq:sigma33}) becomes important, especially for the cooling
of mechanical oscillators with high resonant frequencies
approaching the GHz range. The minimal temperature $T_{m0}$ is directly
proportional to the pumping power in the limit
$\tilde{V}_0\ll 1$, which is the relevant limit in order to
determine the lowest achievable temperature of the mechanical oscillator
cooled by sideband cooling.


\begin{figure}
\includegraphics[width=7.0cm]{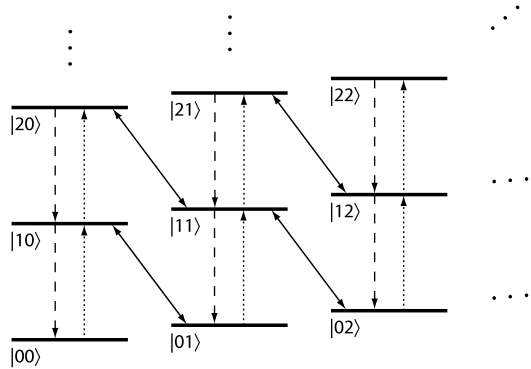}
\caption{Energy-level diagram of a high-frequency resonator and
low-frequency mechanical oscillator with different possible
transitions. The first and second
quantum numbers represent the number of excitations in the
resonator and mechanical oscillator, respectively. The vertical arrows represent the environment-induced
decay in the resonator, and slanted arrows represent
driving-induced transitions. Decay in the mechanical oscillator is
assumed to be negligibly small.}
\label{Fig:EnergyLadder1}
\end{figure}

\section{\label{sec:Quantum} Quantum approach}
In order to achieve the quantum regime of the mechanical
oscillator, the temperature should be lower than the energy of
quantum fluctuations which, together with  Eq.~(\ref{Eq:T_ml}),
imply the inequality
$$\frac{\omega_mT_r}{\omega_r}<T^*_m<\frac{\hbar\omega_m}{k_B}\ .$$
Thus the microwave resonator should be in the quantum regime as
well, and the classical description is no longer valid. Therefore,
we now turn to the analysis of this problem using the quantum
description when the resonator's frequency is higher than its
temperature and the resonator is in its ground state with high
probability. The analysis is also valid if the resonator is substituted
by a two-level system (qubit) as sugested in Ref.~\onlinecite{martin04}.
In this case the cooling limit can be derived in a
transparent manner.

The Hamiltonian that we shall use in our analysis is given by
\begin{equation}
\hat{H} = \omega_r a_r^{\dagger} a_r + \omega_m a_m^{\dagger} a_m
+ \hat{H}_{\rm coupling} + \hat{H}_{\rm drive},
\end{equation}
where $a_r^{\dagger}$ and $a_r$ ($a_m^{\dagger}$ and $a_m$) are,
respectively, the creation and annihilation operators of the
resonator (oscillator). The term $\hat{H}_{\rm coupling}$
represents the oscillator-resonator coupling, and the term
$\hat{H}_{\rm drive}$ represents the driving force. We shall
assume that the last two terms in the Hamiltonian are small: The
smallness of $\hat{H}_{\rm coupling}$ means that the energy
eigenstates will, to a good approximation, be identified with
well-defined excitation numbers in the oscillator and resonator,
while the smallness of $\hat{H}_{\rm drive}$ justifies a
description of the system using time-independent energy levels. In
the following we start by using thermodynamics arguments to derive
an expression for the lower limit on the achievable temperature,
and we later use a master-equation approach to treat the specific
example discussed in Sec.~\ref{sec:Classical}.

We first consider the situation depicted in
Fig.~\ref{Fig:EnergyLadder1}. Each arrow describes a transition
from a state $|i,j\rangle$ to another state $|i',j'\rangle$, where
the meaning of the quantum numbers is explained in
Fig.~\ref{Fig:EnergyLadder1}. We denote the rate at which such a
transition occurs by $W_{|i,j\rangle \rightarrow |i',j'\rangle}$.
In other words, the probability current of the transition is given
by $P_{|i,j\rangle} W_{|i,j\rangle \rightarrow |i',j'\rangle}$,
where $P_{|i,j\rangle}$ is the occupation probability of the state
$|i,j\rangle$. In the steady state, we can write detailed-balance
equations for the occupation probabilities of the different
quantum states in the form
\begin{widetext}
\begin{eqnarray}
0 = \frac{dP_{|i,j\rangle}}{dt} & = & \left( W_{|i+1,j-1\rangle
\rightarrow |i,j\rangle} P_{|i+1,j-1\rangle} - W_{|i,j\rangle
\rightarrow |i+1,j-1\rangle} P_{|i,j\rangle} \right) + \left(
W_{|i-1,j+1\rangle \rightarrow |i,j\rangle} P_{|i-1,j+1\rangle} -
W_{|i,j\rangle \rightarrow |i-1,j+1\rangle} P_{|i,j\rangle}
\right) \nonumber
\\
& & + \left( W_{|i+1,j\rangle \rightarrow |i,j\rangle}
P_{|i+1,j\rangle} - W_{|i,j\rangle \rightarrow |i+1,j\rangle}
P_{|i,j\rangle} \right) + \left( W_{|i-1,j\rangle \rightarrow
|i,j\rangle} P_{|i-1,j\rangle} - W_{|i,j\rangle \rightarrow
|i-1,j\rangle} P_{|i,j\rangle} \right)
\label{Eq:Detailed_balance}
\end{eqnarray}
\end{widetext}
Now we determine some relations among
rates $W$. Let us start with situation when the driving force is
switched off and the resonator is in contact with its surrounding
environment, which is at temperature $T_r$. Assuming that the
environment induces transitions between states that are different
by one photon in the resonator, the rates must obey the
thermal-equilibrium relation
\begin{equation}
\frac{W_{|i,j\rangle \rightarrow |i+1,j\rangle}}{W_{|i+1,j\rangle
\rightarrow |i,j\rangle}} = \exp \left\{ - \,
\frac{\hbar\omega_r}{k_BT_r} \right\}.
\label{Eq:Rates1}
\end{equation}
Note that these transitions do not change the state of the
mechanical oscillator, since without the driving force the
oscillator and resonator are effectively decoupled
($\omega_m\ll\omega_r$). The oscillator is itself in contact with
its environment at temperature $T_m$, but for optimal cooling we
assume that the insulation is good enough that we can completely
neglect environment-induced transitions, i.e.~we have assumed that
$W_{|i,j\rangle \rightarrow |i,j\pm 1\rangle}\rightarrow 0$ in
Fig.~\ref{Fig:EnergyLadder1} and Eq.~(\ref{Eq:Detailed_balance}).
We now assume that the driving force couples states of the form
$|i,j\rangle$ and $|i+1,j-1\rangle$ but does not drive any other
transitions. Since the driving force is a classical one, the
transitions it induces must have equal rates in both directions,
i.e.
\begin{equation}
W_{|i,j\rangle \rightarrow |i+1,j-1\rangle} = W_{|i+1,j-1\rangle
\rightarrow |i,j\rangle}.
\label{Eq:Rates2}
\end{equation}
The reason why there is no Boltzmann factor in
Eq.~(\ref{Eq:Rates2}) is that these transitions are mainly induced
by the classical driving force, and any contributions to their
rates from the thermal environment are negligible.

Using Eqs.~(\ref{Eq:Rates1}) and (\ref{Eq:Rates2}), it is not
difficult to verify that the pairs of terms in
Eq.~(\ref{Eq:Detailed_balance}) all vanish when
\begin{equation}
P_{|i,j\rangle} = \frac{1}{Z} \exp \left\{ -
\frac{(i+j)\hbar\omega_r}{k_BT_r} \right\},
\end{equation}
where $Z$ is the partition function. This steady-state probability
distribution $P_{|i,j\rangle}$ can now be rewritten as
\begin{eqnarray}
P_{|i,j\rangle} & = & \frac{1}{Z_r} \exp \left\{ - \frac{i
\hbar\omega_r}{k_BT_r} \right\} \times \frac{1}{Z_m} \exp \left\{
- \frac{j \hbar\omega_r}{k_BT_r} \right\} \nonumber \\
& = & \frac{1}{Z_r} \exp \left\{ - \frac{i \hbar\omega_r}{k_BT_r}
\right\} \times \frac{1}{Z_m} \exp \left\{ - \frac{j
\hbar\omega_m}{k_BT_m^*} \right\}, \nonumber \\
\
\end{eqnarray}
with
\begin{equation}
\frac{T_m^*}{T_r} = \frac{\omega_m}{\omega_r}.
\label{Eq:T*T}
\end{equation}
Here $Z_r$ and $Z_m$ is partition function of resonator and mechanical
oscillator, respectively.
We therefore find that if the above picture about the allowed
transitions and the relations governing their rates are valid, we
can reach the final temperature $T_m^*$ given by
Eq.~(\ref{Eq:T*T}).

The above derivation suggests an intuitive picture for the cooling
mechanism. The purpose of the driving force is to facilitate the
transfer of excitations between the resonator and oscillator.
Before the driving starts, the low-frequency oscillator has many
more excitations than the high-frequency resonator. Once the
driving starts, the excitation imbalance causes excitations to
start flowing from the oscillator to the resonator. As the number
of excitations in the resonator goes above the thermal-equilibrium
value, excitations start to dissipate from the resonator to the
environment. A steady state is eventually reached with the
resonator in thermal equilibrium with the environment and both the
resonator and the oscillator having the same average number of
excitations (in fact, the resonator and the oscillator will have
the same excitation-number probability distribution). This picture
of the cooling mechanism reveals another point that is generally
not noted in the literature. Although $\Gamma_r$ is desired to be
smaller than $\omega_m$ in order to avoid heating effects, it
should not be too small, because it provides the mechanism by
which excitations are dissipated from the resonator into the
environment. In particular, it must be larger than $\Gamma_m$,
such that the dissipation of excitations is faster than the
heating of the oscillator by its environment.

\begin{figure}
\includegraphics[width=7.0cm]{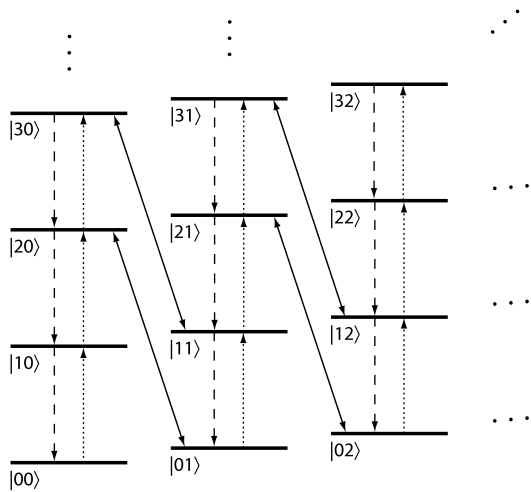}
\caption{Same as in Fig.~\ref{Fig:EnergyLadder1}, but with the
driving force inducing a different type of transitions. The driven
transitions in this case remove one excitation from the oscillator
state and add two excitations to the resonator state, or vice
versa.}
\label{Fig:EnergyLadder2}
\end{figure}
We now consider what would happen if one were able to drive
transitions as shown in Fig.~\ref{Fig:EnergyLadder2}. With optimal
parameters for cooling, one would obtain the minimum temperature
\begin{equation}
\frac{T_m^*}{T_r} = \frac{\omega_m}{2\omega_r}.
\label{Eq:T*T2}
\end{equation}
Note that this situation would require driving the system at a
frequency $2\omega_r-\omega_m$. The argument can be generalized to
obtain any value $n$ in the denominator of Eq.~(\ref{Eq:T*T2}), or
more explicitly
\begin{equation}
\frac{T_m^*}{T_r} = \frac{\omega_m}{n\;\omega_r}.
\label{Eq:T*Tn}
\end{equation}
The question is whether such transitions can be realistically
driven in a given system with a specific form of
resonator-oscillator coupling and a given type of driving force.

In order to determine the feasibility of realizing conditions where
Eq.~(\ref{Eq:T*Tn}) with $n>1$ is the relevant lower limit on the
achievable temperature, we now turn from the above general arguments
to the specific situation considered in Sec.~\ref{sec:Classical}.
Equations (\ref{Eq:LRCxa}),(\ref{Eq:LRCxb}) result from a Hamiltonian of the form
\begin{eqnarray}
\hat{H} &=& \omega_r a_r^{\dagger} a_r + \omega_m a_m^{\dagger}
a_m + g \left(a_r+a_r^{\dagger}\right)^2
\left(a_m+a_m^{\dagger}\right) \nonumber \\
&+& A \cos(\omega_p t + \theta)
\left(a_r+a_r^{\dagger}\right),
\label{Eq:Hamiltonian}
\end{eqnarray}
where $g$ is the oscillator-resonator coupling strength, and $A$
is the amplitude of the driving force. Starting with a simplified
version of the above Hamiltonian that does not contain the last
two terms, we have an energy-level diagram similar to the one
shown in Fig.~\ref{Fig:EnergyLadder1} (without the
induced-transition arrows). The resonator-oscillator coupling term
mixes the different quantum states together in the eigenstates of
the Hamiltonian. This mixing allows the driving term, which would
normally affect only the resonator, to drive transitions that
remove excitations from the oscillator and add excitations to the
resonator, or vice versa. Therefore, in order to determine the
transitions that can be driven by the external force, we need to
evaluate matrix elements of the form $\langle \psi_{i,j}|
(a_r+a_r^{\dagger}) |\psi_{k,l}\rangle$, where we now use the
eigenstates of the Hamiltonian (i.e. slightly modified from the
case of two uncoupled oscillators). To first order in perturbation
theory,
%
\begin{widetext}
\begin{eqnarray}
|\psi_{i,j}\rangle & \approx & |i,j\rangle +
\frac{(2i+1)g}{\omega_m} \left( \sqrt{j} |i,j-1\rangle -
\sqrt{j+1} |i,j+1\rangle \right) \nonumber \\
& & +
\frac{g}{2\omega_r - \omega_m} \left( \sqrt{i(i-1)(j+1)}
|i-2,j+1\rangle - \sqrt{(i+1)(i+2)j} |i+2,j-1\rangle \right).
\nonumber \\
\end{eqnarray}
\end{widetext}
Using the above approximation, we find that
\begin{equation}
\langle \psi_{i,j}| a_r+a_r^{\dagger} |\psi_{i+1,j-1}\rangle \approx -
\frac{2g\sqrt{(i+1)j}}{\omega_m}.
\end{equation}
It is straightforward to see from Eq.~(\ref{Eq:Hamiltonian}) that
$$\langle \psi_{i,j}| a_r+a_r^{\dagger} |\psi_{i+2,j-1}\rangle=0\ .$$
Using numerical calculations we find that
\begin{equation}
\langle \psi_{i,j}| a_r+a_r^{\dagger} |\psi_{i+3,j-1}\rangle \approx -
\frac{12 g^3 \sqrt{(i+1)(i+2)(i+3)j}}
{\left(2\omega_r-\omega_m\right)^3}.
\end{equation}

The above results imply that the driving term can be used to drive
transitions of the form $|i,j\rangle \leftrightarrow
|i+1,j-1\rangle$, which can be used to remove excitations from the
oscillator and add them to the resonator. These transitions
correspond to the picture shown in Fig.~\ref{Fig:EnergyLadder1},
and their resonance frequency is given by
$\omega_p=\omega_r-\omega_m$. The steady-state effective
temperature for the oscillator is given by Eq.~(\ref{Eq:T*T}) when
driving these transitions, assuming that heating effects are
avoided. By driving the system at the frequency
$\omega_p=3\omega_r-\omega_m$, one could in principle drive the
transitions $|i,j\rangle \leftrightarrow |i+3,j-1\rangle$ and
reach a lower minimum temperature. However, the fact that the
corresponding matrix element is proportional to the third power of
the small coupling strength $g$ suggests that this matrix element
will be extremely small for any realistic parameters, hindering
the possibility of utilizing this cooling mechanism.

We now turn to the heating effects that have been neglected above.
We note that the driving term in Eq.~(\ref{Eq:Hamiltonian}) can
also drive transitions of the form $|i,j\rangle \leftrightarrow
|i+1,j+1\rangle$, and the relevant matrix element is given by
\begin{equation}
\langle \psi_{i,j}| a_r+a_r^{\dagger} |\psi_{i+1,j+1}\rangle \approx
\frac{2g\sqrt{(i+1)(j+1)}}{\omega_m}.
\end{equation}
These undesired transitions are induced if either the driving
amplitude $A$ or the resonator's damping rate $\Gamma_r$ is
comparable to or larger than $\omega_m$. If either one or both of
the above conditions are satisfied, the driving force must be
considered within the resonance region of the above transition. As
a result, additional excitation would be steadily pumped into the
system, resulting in a higher temperature than what would be
obtained from the simple picture of transition rates that we have
presented above. The ideal parameters for cooling are therefore
given by $\omega_p=\omega_r-\omega_m$, $A \ll (\omega_r-\omega_p)$
and $\Gamma_r \ll (\omega_r-\omega_p)$; naturally $\Gamma_m$ is
desired to be much smaller than both $\Gamma_r$ and
$A^2/\Gamma_r$, such that the oscillator heating from its contact
with the environment is slower than the cooling it experiences as
a result of the driving. Using a numerical simulation, we shall
see shortly that the above heating effects can be made negligible
with the proper choice of parameters. We should also mention here
that in this section we have not considered the noise in the
driving force, i.e. we have assumed a perfect microwave source.
Such noise would directly heat the resonator, resulting in a
higher base temperature, as discussed in Sec.~\ref{sec:Classical}.

\begin{figure}[!t]
    \psfrag{TeffAA}{$T_{m,r}^*/T$}
\psfrag{timeAAAAA}{$t$ ($2\pi/\omega_m$)}
\includegraphics[width=7.0cm]{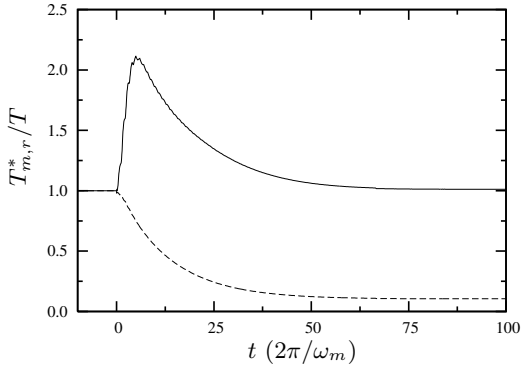}
\caption{Effective temperatures for the resonator (solid line) and
the oscillator (dashed line), relative to the ambient temperature,
as a function of time for the parameters $\omega_m/\omega_r =
0.1$, $g/\omega_r = 0.005$, $k_BT_{m,r}/\hbar\omega_r=0.5$,
$\Gamma_r/(\omega_r/2\pi) = 0.05$, and $\Gamma_m=0$. The driving
field, with amplitude $A/\omega_r = 0.05$ and frequency $\omega_p
= \omega_r-\omega_m$, is turned on at $t=0$.}
\label{Fig:T_t}
\end{figure}

In order to give a concrete example that illustrates the cooling
dynamics, we now turn to a master-equation approach. The density
matrix $\rho$ of the system evolves in time according to the
master equation
\begin{widetext}
\begin{eqnarray}
\frac{d\rho}{dt} & = & - \frac{i}{\hbar} \left[ \hat{H} , \rho
\right] + \left( 1 + \bar{N}_r \right) \Gamma_r \left( a_r \rho
a_r^{\dagger} - \frac{1}{2} a_r^{\dagger} a_r \rho - \frac{1}{2}
\rho a_r^{\dagger} a_r \right) + \bar{N}_r \Gamma_r \left(
a_r^{\dagger} \rho a_r - \frac{1}{2} a_r a_r^{\dagger} \rho -
\frac{1}{2} \rho a_r a_r^{\dagger} \right) \nonumber
\\
& & + \left( 1 + \bar{N}_m \right) \Gamma_m \left( a_m \rho
a_m^{\dagger} - \frac{1}{2} a_m^{\dagger} a_m \rho - \frac{1}{2}
\rho a_m^{\dagger} a_m \right) + \bar{N}_m \Gamma_m \left(
a_m^{\dagger} \rho a_m - \frac{1}{2} a_m a_m^{\dagger} \rho -
\frac{1}{2} \rho a_m a_m^{\dagger} \right),
\label{Eq:Master_equation}
\end{eqnarray}
\end{widetext}
where
\begin{equation}
\bar{N}_r = \frac{1}{e^{\hbar\omega_r/k_BT} - 1}
\end{equation}
and similarly for $\bar{N}_m$. The coefficients $\Gamma_r$ and
$\Gamma_m$ are decay rates for the resonator and oscillator,
respectively.

An example illustrating the dynamics of cooling the mechanical
oscillator by the microwave resonator is shown in
Fig.~\ref{Fig:T_t}. The results were obtained by numerically
solving Eq.~(\ref{Eq:Master_equation}) using the Hamiltonian in
Eq.~(\ref{Eq:Hamiltonian}).
%
%
The effective temperatures of the oscillator and resonator are
obtained by calculating their respective entropies from their
reduced density matrices ($S=-{\rm Trace} \left\{ \rho \log \rho
\right\}$) and fitting these values to the temperature-entropy
relation for a harmonic oscillator. The initial heating of the
resonator is a result of the transfer of excitations from the
oscillator to the resonator. For large $t$, the system reaches a
steady state where the ratio between the effective temperatures of
the oscillator and the resonator is approximately equal to
$\omega_m/\omega_r$.

\section{\label{sec:Conclusions} Conclusions}
We have shown that both the classical and the quantum
treatment give the same final result: the cooling factor
$T^*_m/T_r$ is limited by the ratio $\omega_m/\omega_r$. This
lower limit for the cooling becomes crucial for rf and
microwave resonators pumped by a real (noisy) microwave source
since their effective temperature $T_r$ is usually much larger
than the ambient temperature. We should also emphasize that our
results apply, with minor modifications, to other types of
coolers, e.g. a cooper-pair box.

\acknowledgments
This work was supported in part by the National Security Agency
(NSA), the Laboratory for Physical Sciences (LPS), the Army
Research Office (ARO), the National Science Foundation (NSF) grant
No.~EIA-0130383 and the Japan Society for the Promotion of Science
Core-To-Core (JSPS CTC) program. M.G. was partially supported by
Grants VEGA 1/0096/08, APVT-51-016604 and Center of Excellence of the Slovak Academy of Sciences (CENG).


\end{document}